\documentclass[a4paper,12pt]{article}
\usepackage{a4}
\usepackage{epsfig}
\usepackage{latexsym, amssymb} 
\usepackage{amsmath}
\usepackage{psfrag}
\usepackage{bm}
\usepackage{feynmp}
\usepackage{rotating}
\usepackage{hyperref}
\usepackage[utf8]{inputenc}
\hypersetup{
    unicode=false,          
    pdftoolbar=true,        
    pdfmenubar=true,        
    pdffitwindow=false,     
    pdfstartview={FitH},    
    pdftitle={My title},    
    pdfauthor={Author},     
    pdfsubject={Subject},   
    pdfcreator={Creator},   
    pdfproducer={Producer}, 
    pdfkeywords={keyword1} {key2} {key3}, 
    pdfnewwindow=true,      
    colorlinks=true,        
    linkcolor=red,          
    citecolor=blue,         
    filecolor=magenta,      
    urlcolor=green,         
    linktocpage=true
}

\def\br{\begin{eqnarray}}
\def\er{\end{eqnarray}}
\def\benu{\begin{enumerate}}
\def\eenu{\end{enumerate}}
\def\nn{\nonumber}

\def\l{\left}
\def\r{\right}

\def\beq{\begin{equation}}
\def\eeq{\end{equation}}
\def\apj{Astrophys.\ J.}

\def\mnras{Mon.\ Not.\ Roy.\ Astron.\ Soc.}

\def\prd{Phys.\ Rev.\  D}

\def\bsi{b_{\rm S}^{(1)} }
\def\brsi{b_{\rm R[S]}^{(1)} }
\def\brsii{b_{\rm R[S]}^{(2)} }

\def\bsii{b_{\rm S}^{(2)} }
\def\brii{b_{\rm R}^{(2)} }
\def\bri{b_{\rm R}^{(1)} }

\def\vk{{\bf k}}
\def\vka{{\bf k}_{1}}
\def\vkb{{\bf k}_{2}}
\def\vkc{{\bf k}_{3}}
\def\ka{k_{1}}
\def\kb{k_{2}}
\def\kc{k_{3}}

\def\cB{{\cal B}}
\def\fnl{f_{_{\rm NL}}}

\def\F{\mathcal{F}}

\begin{document}
\title{Probing primordial non-Gaussianity:  The 3D Bispectrum of Ly-$\alpha$ forest and the redshifted 21-cm signal from the post reionization epoch}
\author{Tapomoy Guha Sarkar\footnote{tapomoy@bits-pilani.ac.in}\\
Department of Physics,  Birla Institute of Technology and Science,\\ Pilani, Rajasthan, India.\\
Dhiraj Kumar Hazra\footnote{Current affiliation: Asia Pacific Center for Theoretical Physics, Pohang, 
Gyeongbuk 790-784, Korea.
\vskip 2pt
E-mail:~dhiraj@apctp.org}\\
Harish-Chandra Research Institute Chhatnag Road,\\ Jhunsi, Allahabad~211019, India}
\maketitle

\begin{abstract}
We explore possibility of using the three dimensional bispectra of the
Ly-$\alpha$ forest and the redshifted 21-cm signal from the
post-reionization epoch to constrain primordial non-Gaussianity. Both
these fields map out the large scale distribution of neutral hydrogen
and maybe treated as tracers of the underlying dark matter field. We
first present the general formalism for the auto and cross bispectrum
of two arbitrary three dimensional biased tracers and then apply it to
the specific case.  We have modeled the 3D Ly-$\alpha$ transmitted
flux field as a continuous tracer sampled along 1D skewers which
corresponds to quasars sight lines. For the post reionization 21-cm
signal we have used a linear bias model.  We use a Fisher matrix
analysis to present the first prediction for bounds on $\fnl$ and the
other bias parameters using the three dimensional 21-cm bispectrum and
other cross bispectra.  The bounds on $\fnl$ depend on the survey
volume, and the various observational noises. We have considered a
BOSS like Ly-$\alpha$ survey where the average number density of
quasars $ \bar n = 10^{-3}~{\rm Mpc^{-2}}$ and the spectra are
measured at a 2-$\sigma$ level.  For the 21-cm signal we have
considered a 4000 $\rm hrs $ observation with a futuristic SKA like
radio array. We find that bounds on $f_{\rm NL}$ obtained in our
analysis ($ 6 \le \Delta f_{\rm NL} \le 65 $) is competitive with CMBR
and galaxy surveys and may prove to be an important alternative
approach towards constraining primordial physics using future data
sets. Further, we have presented a hierarchy of power of the
bispectrum-estimators towards detecting the $\fnl$.  Given the quality
of the data sets, one may use this method to optimally choose the
right estimator and thereby provide better constraints on $f_{\rm
  NL}$. We also find that by combining the various cross-bispectrum
estimators it is possible to constrain $f_{\rm NL} $ at a level $
\Delta f_{\rm NL} \sim 4.7$.  For the {\it equilateral} and {\it
    orthogonal} template we obtain $\Delta \fnl^{equ} \sim 17$ and $
  \Delta \fnl^{orth} \sim 13 $ respectively for the combined estimator. This shall
be important in the quest towards understanding the mechanism behind
the generation of primordial perturbations.

\end{abstract}
\tableofcontents
\section{Introduction}

In the study of inflationary cosmology, probing the deviations from
Gaussian initial conditions offer valuable insights to our
understanding of the mechanisms that generated primordial
fluctuations. Whereas, departures from non-Gaussianity is very small
for the standard single inflaton models \cite{maldacena-2003}, a wide
class of theories predict a moderate to high
non-Gaussianity \cite{bartolo-2004}. Measuring the degree of primordial
non-Gaussianity hence, directly allows us to narrow down the range of
viable inflation models. The bispectrum or the three point correlation
function of the density fluctuations is a standard quantifier of
non-Gaussianity and the bispectrum of the CMBR temperature
anisotropies and large scale structure have been extensively studied in this 
regard \cite{wmap-7a,wmap-7b, wmap-7c, bern2010}.   Study of large scale structure 
using galaxy redshift surveys have also been proposed as means to constrain non-Gaussianity \cite{socci-sefu, sefu2007}. A recent work has also proposed the use of three
dimensional Ly-$\alpha$ forest to measure primordial non-Gaussianity
\cite{hazrasarkar-2012}.  The bispectrum of any good tracer of the
underlying density field is, however a potential probe of primordial
non-Gaussianity.

The neutral hydrogen (HI) distribution in the the post reionization
IGM is a powerful cosmological probe of the low redshift universe. The
complex astrophysical processes that dictate the HI distribution
during the epoch of reionization, becomes largely irrelevant at
redshifts $z < 6$. Here, two astrophysical systems are of
observational interest.  The dense and self shielded Damped Lyman
$\alpha$ (DLA) systems \cite{wolfe05} house bulk of the HI and is the
dominant source of the 21-cm signal \cite{hirev1}(seen in emission).
 The HI density fluctuations in the
  predominantly ionized IGM, on the contrary is responsible for the
  distinct absorption features - the Ly-$\alpha$ forest
  \cite{rauch1998}, in the spectrum of back ground quasars.  On large
  cosmological scales both the Ly-$\alpha$ forest and the redshifted
  21-cm signal are believed to be biased tracers of the underlying
  dark matter distribution.

 We note that the 21-cm signal from individual DLA clouds is extremely
  small. There, however maybe some enhancement due to gravitational
  lensing \cite{saini}, but for most purposes it is reasonable to look at
  the large scale diffuse distribution of the signal which forms a
  background in radio observations. This low resolution mapping of the
  21-cm sky over large volumes at redshifts $z < 3$ is called
  `Intensity Mapping' \cite{chang2008, wyitheloeb}. The large number density of DLAs make
the Poisson noise arising from discrete sampling almost negligible \cite{wyitheloeb}.

The possibility of measuring the cross correlation power spectrum of
the 21-cm signal and Ly-$\alpha$ forest has been proposed \cite{tgs5}
as a means to bypass some of the observational issues. The two
signals being tracers of the underlying large scale structure are
expected to be correlated on large scales. However, foregrounds and
other systematics are believed to be uncorrelated between the two
independent observations. Hence, the cross correlation power spectrum
signal if detected will clearly ascertain its cosmological origin.  It
is only natural to generalize the notion of cross power spectrum to
cross-bispectrum of the two fields. The cross-bispectrum is an obvious
byproduct of the two data sets that are used to measure the auto
correlations.  Further, an advantage of the cross-correlation study is
that, in it, the demerits of one of the data sets gets partly
compensated by the merits of the other.

The auto correlations of both the 21-cm maps \cite{joudaki11} and the
Ly-$\alpha$ forest \cite{hazrasarkar-2012} towards a measurement of
primordial non-Gaussianity has been independently studied.  In this
paper we investigate the three dimensional cross-bispectrum of these
cosmological fields. We set up the general formalism for calculating
the cross-bispectrum of two arbitrary low redshift tracers of the
underlying matter distribution. The general formalism is then applied
to estimate the cross-bispectrum of 21-cm brightness temperature and
Ly-$\alpha$ transmitted flux fields.  We use a Fisher matrix analysis
to investigate the possibility of constraining the non-Gaussianity
parameter $f_{\rm NL}$ from the cross-bispectrum signal for a range of
observational parameters. A comparison between the various auto and
cross bispectra is also presented.  We note that this is the first direct
investigation of primordial non-Gaussianity using bispectrum of the
entire three dimensional information contained in the distribution of
low redshift neutral hydrogen.

\section{Formalism}
 On sub-horizon scales the matter overdensity field
$\Delta_{\vk}$ in Fourier space is related to the primordial
gravitational potential as 
\beq
 \Delta_{\vk} (z) = {\mathcal{M}}(k, z) 
\Phi^{\rm prim}_{\vk} = -\frac{3}{5} \frac{k^2 T(k)}{\Omega_m H_0^2}
D_{+}(z) \Phi^{\rm prim}_{\vk},
\eeq
 where $T(k)$ denotes the matter
transfer function and $D_{+}(z)$ is the growing mode of density
fluctuations.  The statistical properties of $\Delta_{\vk}$ are
quantified through the $n-$point correlations. The first two
non-trivial of these are the power spectrum and bispectrum, defined
respectively as 
\begin{eqnarray}
 \langle \Delta_{\vka}\Delta_{\vkb} \rangle &=&
\delta_D(\vka + \vkb) P(\ka) \nonumber \\
\langle \Delta_{\vka}
\Delta_{\vkb} \Delta_{\vkc}\rangle &=& \delta_D(\vka + \vkb + \vkc)
B(\ka, \kb, \kc).  
\end{eqnarray}
The assume that the primordial gravitational potential $\Phi^{\rm prim} = \phi_G$  may be written
as 
\beq
 \Phi^{\rm prim} = \phi_G + \frac{f_{\rm NL}}{c^2} \left( \phi_G ^2 -
\langle \phi_G ^2 \rangle \right),
\eeq 
 where, $ \phi_G $ is a Gaussian random
field and non-Gaussianity is quantified using a single parameter $
f_{\rm NL}$.  The linear power spectrum of a sufficiently smoothed
density field is given by 
\beq
P^{\rm L} (k)= \frac{9}{25} \frac{k^4
  T^2(k)}{\Omega_m^2 H_0^4} D_{+}^2(z) P_{\Phi}^{\rm prim},
\eeq 
where $P_{\Phi}^{\rm prim}$ denotes the primordial power spectrum of the
gravitational potential such that $ P_{\Phi}^{\rm prim} = P_{\phi {G}}
+ {\cal{O}} (\fnl ^2)$.  The power spectrum $ P_{\phi {G}} $ arising
from the Gaussian field $\phi_G$ does not exhibit exotic features and
is scale invariant.  In a linear theory the bispectrum arising from
non-Gaussianity in the primordial matter field is given by
\beq
{B^{\rm
    L}}_{123} = {\cal{M}}(\ka){\cal{M}}(\kb){\cal{M}}(\kc) {B_{\phi
    G}}_{123}. 
\eeq
Here the notation ${123} \equiv (\ka, \kb, \kc)$ and
we have used the form of $ {B_{\phi G}}_{123}$ from earlier works \cite{bslss, hazrasarkar-2012}.
 \br {B_{\phi G}}_{123}= \frac{2 \fnl}{c^2}\left [ P_{\phi 
    G}(\ka) P_{\phi G} (\kb) + \rm {cyc}\right] + {\cal{O}} (\fnl ^3).
 \er 
Here we note that we are using the {\it local} template in the definition of $\fnl$. This choice of template makes the contribution to bispectrum predominantly coming from squeezed triangles. 
The {\it equilateral } and {\it orthogonal} templates upto linear order in $\fnl$ are defined as \cite{sefu2007, socci}
\br
{B_{\phi G}}_{123} = \frac{6 \fnl^{equ}}{c^2}\left [\left( - P_{\phi 
    G}(\ka) P_{\phi G} (\kb)  + 2  ~{\rm cyc}  \right) -  \left( 2 P_{\phi  G}(\ka)^{\frac{2}{3}}P_{\phi  G}(\kb)^{\frac{2}{3}}P_{\phi  G}(\kc)^{\frac{2}{3}} \right)\right . \nonumber \\  \left. +  \left ( P_{\phi  G}(\ka)^{\frac{1}{3}} P_{\phi  G}(\kb)^{\frac{2}{3}}P_{\phi  G}(\kc)  + 5  ~{\rm cyc} \right)\right].
 \er 

\br
{B_{\phi G}}_{123}= \frac{6 \fnl^{orth}}{c^2}\left [- \left(3 P_{\phi  G}(\ka) P_{\phi G} (\kb)  + 2  ~{\rm cyc} \right) - \left ( 8 P_{\phi  G}(\ka)^{\frac{2}{3}}P_{\phi  G}(\kb)^{\frac{2}{3}}P_{\phi  G}(\kc)^{\frac{2}{3}} \right)\right . \nonumber \\  
\left. + \left(  3 P_{\phi  G}(\ka)^{\frac{1}{3}} P_{\phi  G}(\kb)^{\frac{2}{3}}P_{\phi  G}(\kc)  + 5  ~{\rm cyc} \right)\right].
\er 
Henceforth we shall use $\fnl$ without any superscript to denote the  {\it local} template.

The measurement of $n-$point functions of the matter density field
is usually implicit.  One actually measures the statistical properties
of the low redshift biased tracers. At lower redshifts one needs to
incorporate non-linear structure formation caused by gravitational
instability. This induces additional non-Gaussianity in addition to
the contribution from primordial sources.  The additional contribution
to the matter bispectrum, $ B^{\rm NL}_{123}$ \cite{bslss, hazrasarkar-2012} is obtained from the second
order perturbation theory. Hence,
\beq B^{\rm NL}_{123} = 2 F_2(\vka, \vkb)
P(\ka) P(\kb) + \rm{cyc},   \eeq
where $ F_2(\vka, \vkb)$,   adopted from \cite{bslss} is given by 
\beq
F_2(\vka, \vkb) = \frac{5}{7} + \frac{\hat{k_1} \cdot \hat{k_2}}{2} \left(
\frac{k_1}{k_2} + \frac{k_2}{k_1} \right) + \frac{2}{7} \, {(\hat{k_1} \cdot \hat{k_2})}^2.
\eeq
The total matter bispectrum is hence given by the 
sum
\beq
B_{123} = B^{\rm L}_{123} +
B^{\rm NL}_{123},
\eeq 
where a  possible contribution
from the primordial trispectrum has been ignored. 

Let us consider tracer fields denoted by $R$ and $S$ which  are related to 
the underlying matter overdensity field as
\begin{equation}
R(\vk) \left [ S(\vk) \right] = {\brsi}  \Delta (\vk) +  \frac{\brsii}{2}  \Delta {(\vk)}^2. ~~~ 
\label{eq:tracer}
\end{equation} 

We are interested in the statistical properties of these tracers.
The cross-correlation power spectrum  $P_{RS}$ for the tracer fields and bispectra $\cB_{RRS}$,  in 
the leading order are related to the matter power spectrum and the bispectrum as,
\begin{eqnarray}
P_{RS}(k)&=& \bri \bsi P(k)\nn \\ 
\cB_{RSS}(k_1,k_2,k_3)&=& \frac{\bri {\bsi}^2}{3} \left[ B_{123} +  B_{231} +  B_{312} \right ] +
\frac{1}{3}  \left[ 2 \bri \bsi \bsii ( P(k_1)P(k_2) + \rm cyc.)  \right.  \nonumber \\
&+& \left. \brii  \bsi  \bsi  ( P(k_1)P(k_2) + \rm cyc.) \right ]. 
\label{eq:signal}
\end{eqnarray}
We note that for $\rm R = S$ we have the auto-correlation power spectrum and bispectrum.
Following the formalism in \cite{bslss} we define the generalized bispectra estimators $
\hat \cB_{\epsilon}$ with $ \epsilon = R, RSS$ as 
\br
\hat{\cB}_{\epsilon} = \frac{V_f}{V_{123}} \int_{\ka} d^3 {\bf q}_1
\int_{\kb} d^3 {\bf q}_2 \int_{\kc} d^3 {\bf q}_3 \delta_D( {\bf q}_{123})
\times ~ \Pi_{\epsilon} ( \ka, \kb, \kc).
\er 
Here $ {\bf
  q}_{123} = {\bf q}_{1} + {\bf q}_{2} + {\bf{ q}}_{3} $, ${V_f} = (2
\pi)^3/V$ where $V$ is the survey volume, and $V_{123}$ is given by,

\beq 
V_{123} = \int_{\ka} d^3
   {\bf q}_1 \int_{\kb} d^3 {\bf q}_2 \int_{\kc} d^3 {\bf q}_3
   \delta_D( {\bf q}_{123}),
   \eeq
   
   and the integrals are performed over the
   $q_i-$ intervals $\left (k_i - \frac{\delta k}{2} , k_i +
   \frac{\delta k}{2}\right )$. The integrands are given by  $\Pi_{RSS}
   = (1/3)[ \Delta_{R}^o(\ka)\Delta_{S}^o(\kb)\Delta_{S}^{o} (\kc) + {\rm cyc.}]$. The
   `observed' fields $\Delta_{S}^o(k_i)$ or $\Delta_{R}^o(k_i)$ in
   Fourier space contain noise contributions and are in general
   different from the respective cosmological fields $\Delta_{S}(k_i)$
   and $\Delta_{R}(k_i)$.  The estimators defined above are unbiased
   and their variances can be calculated using the relation
   $\Delta\hat{\cB_{\epsilon}}^2 = \langle \hat{\cB_{\epsilon}}^2
   \rangle - \langle \hat{\cB_{\epsilon}} \rangle ^2$. In the leading
   order this yields the following
 \begin{eqnarray}
{\Delta \hat \cB_{RSS}}^2(k_1,k_2,k_3) &=& \frac{V_f}{9 V_{123}}
[ t \left ( P_{\rm R}^{\rm Tot}(\ka) P_{\rm S}^{\rm Tot}(\kb)
  P_{\rm S}^{\rm Tot}(\kc) + {\rm cyc.}\right)\nn\\
  &+& 2t \left( P_{\rm
    S}^{\rm Tot}(\ka) P_{\rm RS}^{\rm Tot}(\kb) P_{\rm RS}^{\rm
    Tot}(\kc)+ {\rm cyc.}\right)].
\label{eq:var}
\end{eqnarray}
As earlier, $R = S$ gives the variance for the auto correlation bispectrum.
The combinatorial factor $ t = 6, 1$ for equilateral and scalene
triangles respectively and $P^{\rm Tot}$ includes any additional noise
power spectra along with the cosmological power spectrum.
Armed with the results in equation (\ref{eq:signal}) and equation 
(\ref{eq:var}), we shall apply the general formalism to the two tracer
fields of interest namely the Ly-$\alpha$ forest spectra and
redshifted 21-cm signal, both of which are related to the neutral
hydrogen distribution.  

Observations indicate that, following the epoch of reionization, bulk of
the neutral gas is contained in the self shielded DLA systems \cite{wolfe05}. The
collective 21-cm emission from these clouds is known to form a diffused background
in low frequency radio observations. We use $\delta_{T}$ to denote the
fluctuations in the brightness temperature of redshifted 21-cm
radiation. We are interested in the redshift range $ z < 3.5$ whereby
it is reasonable to assume that $\delta_T$ is a biased tracer of the
underlying dark matter distribution. The numerical
simulations of the post reionization 21 -cm signal \cite{bagla, tgs2011}, show that a constant scale
independent bias model is adequate to describe the large scale
distribution of the 21-cm signal at the relevant redshifts.

The transmitted flux $\F$ through the Ly-$\alpha$ forest may be
modeled by assuming that the gas traces the underlying dark matter
distribution \cite{psbs} except on small scales where pressure plays
an important role.  Further, it is believed
that photo-ionization
equilibrium that maintains the neutral fraction,  also leads to a power
law temperature-density relation \cite{tempdens}.  On large scales it
is reasonable to assume that the fluctuations in the transmitted flux
$\delta_{\F} = \left (\F/ \bar{\F} - 1 \right )$ traces the dark
matter field, with the implicit assumption that the Ly-$\alpha$ forest
spectrum has been smoothed over some suitably large length scale.

 With all the assumptions discussed above we may relate fluctuations
 in Fourier space for both the 21-cm temperature and Ly-$\alpha$
 transmitted flux to the dark matter fluctuations as in  equation
 (\ref{eq:tracer}).  Note, that this model depends on $4$ parameters $({b_{\rm
     T}^{(1)} }, {b_{\rm T}^{(2)} }, {b_{\F}^{(1)} }, {b_{\F}^{(2)} }
 )$ apart from the large scale dark matter distribution and the underlying cosmological model.  For the
 21-cm brightness temperature field we have (for the WMAP7
 cosmological parameters) $ {b_{\rm T}^{(1)} } = 4.0 \, {\rm {mK}} \,
 b_{21} \, {\bar{x}_{\rm HI}}(1 + z)^2 {H_0}/{H(z)} $, where
 ${\bar{x}_{\rm HI}}$ is the mean neutral fraction.  The neutral
 hydrogen fraction is believed to be a constant with a value $
 {\bar{x}_{\rm HI}} = 2.45 \times 10^{-2}$ using $\Omega_{gas} \sim
 10^{-3}$ \cite{xhibar}.  The quantity $b_{21}(k,
 z)$ is studied in numerical simulations \cite{bagla, tgs2011}. We
 adopt a constant bias $b_{21} = 2$ at our fiducial redshift $ z =
 2.5$ from these simulation results.  For the Ly-$\alpha$ forest, we
 use an approximate $ {b_{\F}^{(1)} } \approx -0.13 $ from the
 numerical simulations of Ly-$\alpha$ forest\cite{mcd03}.  We note
 however that these numbers suffer from large uncertainties owing to our lack of a complete
 model of the IGM.

A Ly-$\alpha$ survey usually covers a large volume, as compared to a single 
field of view redshifted 21-cm observation. The cross-correlation can however be computed 
only in the region of overlap between the two fields.

  For the Ly-$\alpha$ forest flux measurements the observed flux
  fluctuations in Fourier space is given by 
\beq
  {\Delta}_{{\F}}^o({\bf{k}}) = \tilde{\rho}({\bf{k}}) \otimes
  {\Delta_{\F}}({\bf{k}}) + \Delta_{\F \rm noise}({\bf{k}}),
\eeq  where
  $\tilde{\rho}$ is the sampling window function in Fourier space, and
  $\Delta_{\F \rm noise}({\bf{k}})$ denotes a possible noise term. The
  function takes care of the discrete sampling of the skewers
  corresponding to the quasar lines of sight.
In the variance calculation using equation (\ref{eq:var}) we use
$P_{\F}^{\rm Tot}(k)$ as the total power spectrum of Ly-$\alpha$ flux
given by \cite{mcdeisen} \beq P_{\F}^{\rm Tot}({\bf k} ) = P_{\F}({\bf k}) + P^{\rm
  {1D}}_{\F}(k_{\parallel}) P_{W} + N_{\F},
\label{eq:totalps}
\eeq 
where $ P_{\F}({\bf k}) = {b_{\F}^{(1)} }^2 P( {\bf k})$, the
quantity $P^{\rm 1D}_{\F}(k_\parallel)$ is the usual 1D flux power
spectrum \cite{psbs} corresponding to individual spectra given by
\beq
P^{\rm {1D}}_{\F}(k_{\parallel}) = (2\pi)^{-2} \int d^2{\bf
  k}_{\perp} P_{\F}({\bf k})
\eeq
 and $ P_{W} $ denotes the power spectrum
of the window function. The quantity $ N^{}_{\F}$ denotes the
effective noise power spectra for the Ly-$\alpha$ observations.  The
`aliasing' term $ P^{\rm {1D}}_{\F}(k_{\parallel}) P_{W}$ quantifies
the discreteness of the 1D Ly-$\alpha$ skewers.  If the Ly-$\alpha$
spectra are measured with a sufficiently high SNR, it maybe shown that
each quasar line of sight may actually be given the same weight and $
P_{W} = \frac{1}{\bar{n}}$, where $\bar{n}$ is the 2D density of
quasars ($\bar{n} = N_{Q}/ {\mathcal{A}}$, where ${\mathcal{A}}$ is
the area of the observed field of view) \cite{multiplelos}. We assume
that the variance $\sigma^2_{{\F} N}$ of the pixel noise contribution
to $\delta_{\F}$ is the same across all the quasar spectra. It is related to the signal to noise ratio on the
continuum and depends on the size of the pixel. Hence we
have $N_{\F} = \sigma^{2}_{\F N}/\bar{n}$ for its the  noise power
spectrum.
\beq
N_{\F} = \frac{1}{ \bar n} \langle \bar \F \rangle ^{-2} [S/N]_{\Delta x}^{-2} (\Delta x
/ 1 ~{\rm Mpc}).
\eeq
 We shall henceforth refer to this noise for $S/N$ in $1\mathring{A}$ 
pixels. In arriving at equation (\ref{eq:totalps}) the effect of
quasar clustering \cite{myers} has been ignored with the assumption
that dominant contribution to noise comes from the Poisson term for
Quasar surveys with realistic values of $\bar n$.

For the 21-cm observations one may ignore the discrete nature of the
DLA sources. We consider a radio-interferometric measurement of the
redshifted 21-cm signal with the total signal is given by \beq P^{\rm
  Tot}_{\rm T} = P_{\rm T}({\bf k}) + N_{\rm T}. \eeq 

The cosmological component of the signal is 
$ P_{\rm T}({\bf k}) = (b_{\rm T}^{(1)})^2 P({\bf k})$ and the noise power
spectrum  \cite{mcquinn} $N_{\rm T}$ at an observed frequency $\nu = 1420/(1 + z) {\rm
  MHz}$ can be calculated using the relation  
\beq
N_{\rm T} (k,\nu)=\frac{T_{sys}^2}{B t_0} \left (
\frac{\lambda^2}{A_e} \right)^2 \frac{r_{\nu}^2 l}{n_b(U,\nu)}.
\eeq
Here $T_{sys}$ denotes the sky dominated system temperature, $B$ is the observation
bandwidth, $t_0$ is the total observation time, $r_{\nu}$ is the
comoving distance to the redshift $z$,  $l$ is the comoving length
corresponding to the bandwidth $B$,  $n_b(U,\nu)$ is the number density
of baseline $U$, where $U=k_{\perp}r_{\nu}/2 \pi$, and $A_e$ is the
effective collecting area for each individual antenna. We may write
$n_b(U,\nu) ={N(N-1)}f_{2D}(U,\nu)/2$ where $N$ is the total
number of antenna in the radio array and $f_{2D}(U,\nu)$ is the
normalized baseline distribution function \cite{datta07, petrovic}.

We use a Fisher matrix analysis to constrain $f_{\rm NL} $ from the
observables of relevance here, namely the auto and cross bispectra $\cB$ 
(the index $\epsilon$ is dropped for brevity).
The Fisher matrix for a set of model parameters $p_i$ is given by 
\beq
F_{ij} = \sum_{\rm RS}\sum_{\ka = k_{min}}^{k_{max}} \sum_{\kb = k_{min}}^{\ka}
\sum_{\kc = \tilde{k}_{min}}^{\kb} \frac{1}{\Delta{\cB}_{\rm RSS}^{2}}~
\frac{\partial{\cB}_{\rm RSS}}{\partial p_i}~\frac{\partial{\cB}_{\rm RSS}}{\partial p_j},
\eeq 
where $ \tilde{k}_{min} = {\bf max}(k_{min},
|\ka - \kb|)$ and the summations are performed using $\delta k =
k_{min}$. Assuming that the likelihood functions for $p_i$ are
Gaussian distributed, the corresponding minimum errors in $p_i$ are
limited by the Cramer-Rao bound $\sigma_i^2 = F^{-1}_{ii}$.  We have
modeled the Ly-$\alpha$ and 21-cm signals using a set of bias
parameters and the non-Gaussianity parameter so that $p_i = (f_{\rm NL},  {b_{\rm T}^{(1)} }, {b_{\rm T}^{(2)} },
{b_{\F}^{(1)} }, {b_{\F}^{(2)} } )$.  We shall  use the Fisher matrix
to investigate the power of a Ly-$\alpha$ survey and 21-cm observation
to constrain these parameters.
\section{Results and Discussion}
\noindent
\begin{figure}[!htb]
\begin{center}
\resizebox{140pt}{90pt}{\includegraphics{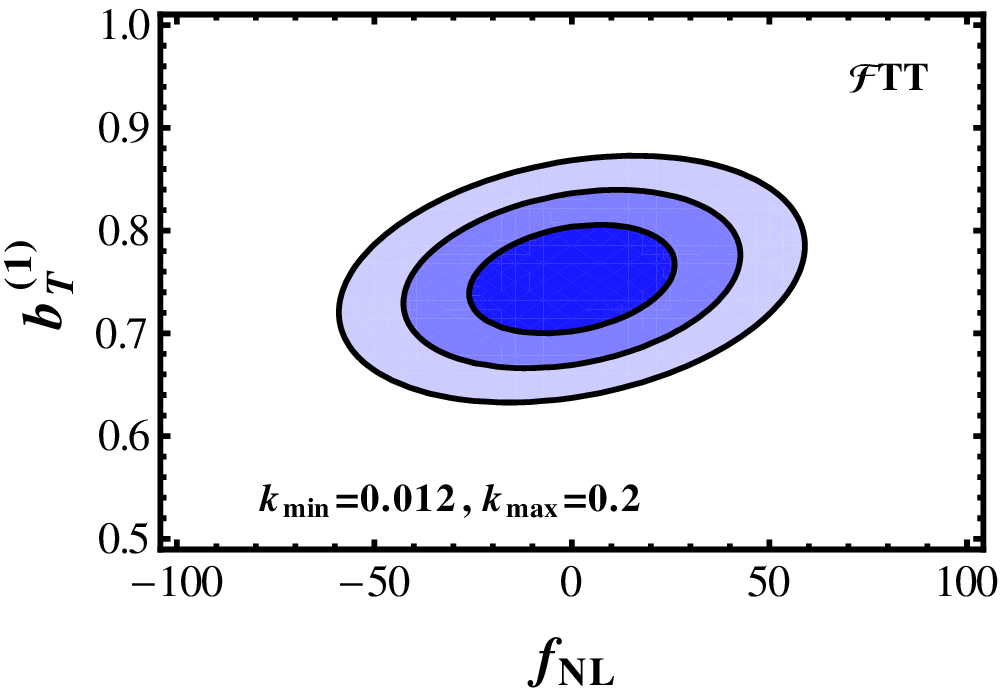}}
\resizebox{140pt}{90pt}{\includegraphics{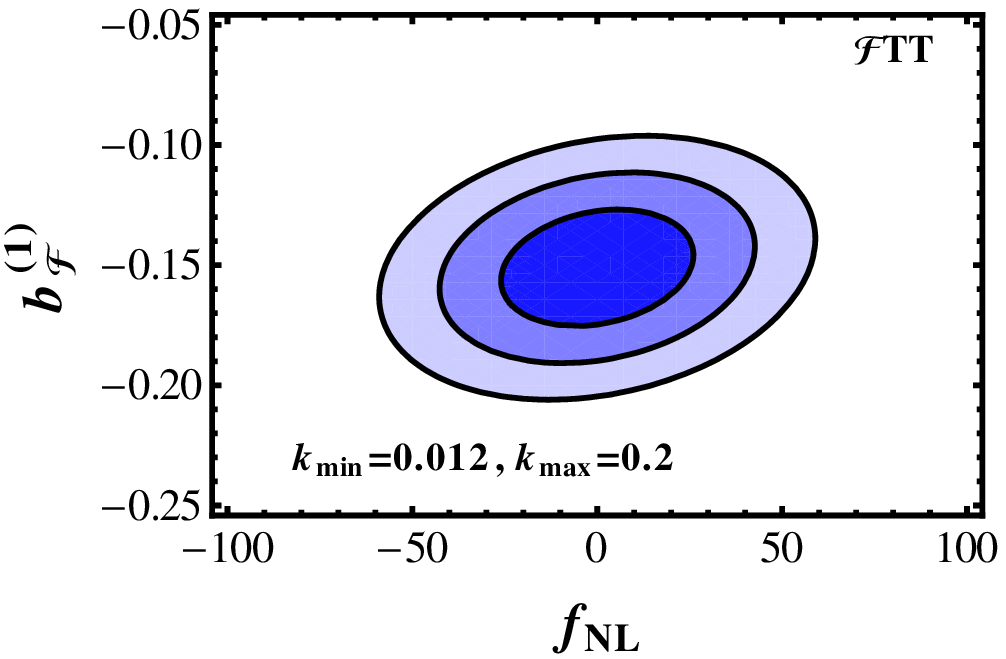}}
\resizebox{140pt}{90pt}{\includegraphics{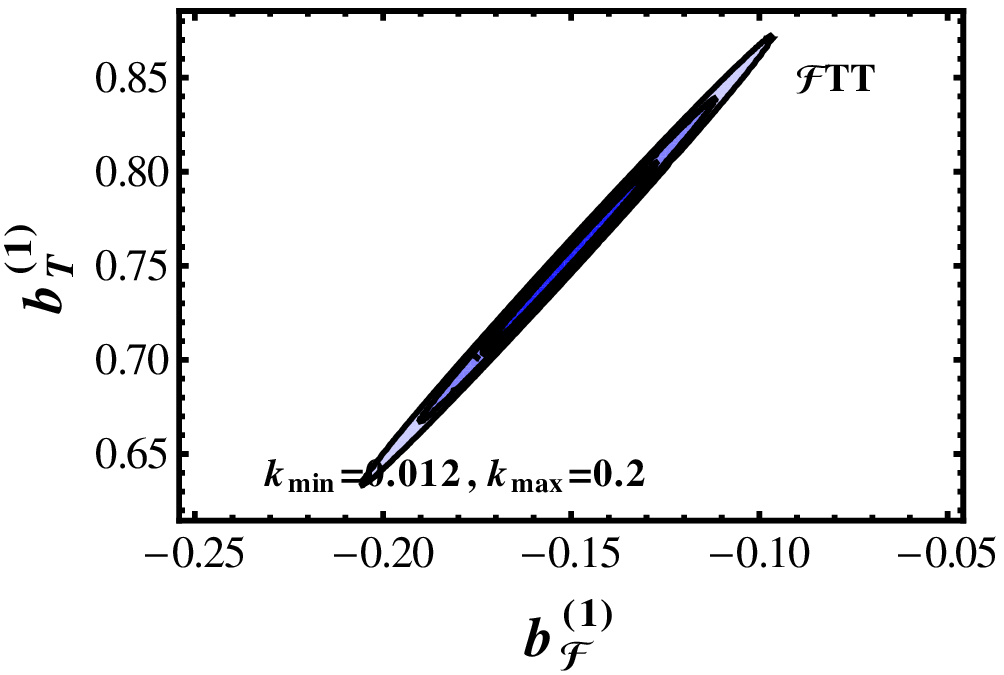}}

\resizebox{140pt}{90pt}{\includegraphics{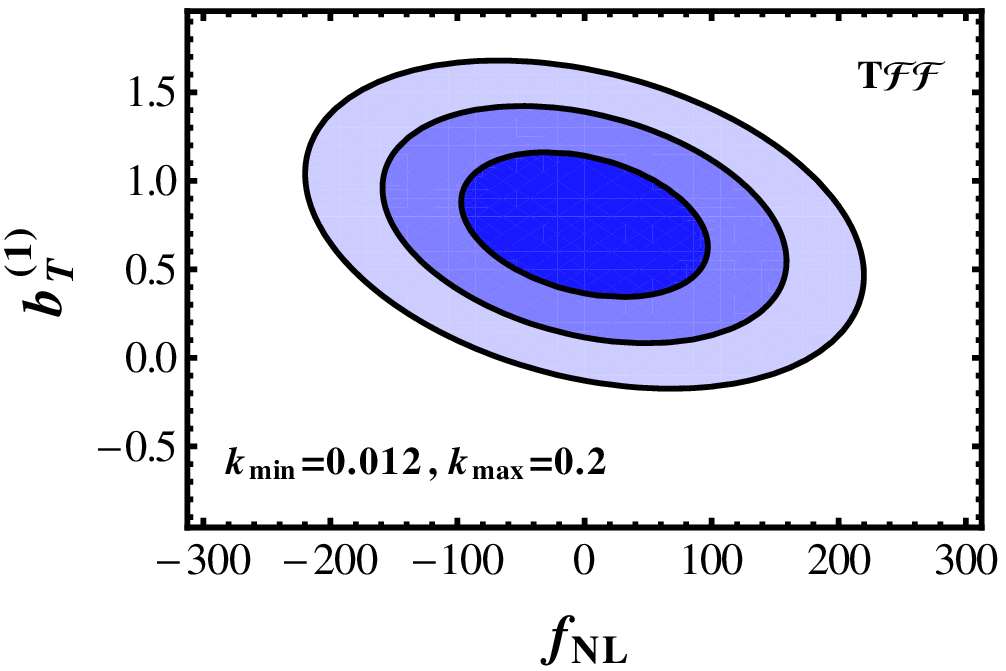}}
\resizebox{140pt}{90pt}{\includegraphics{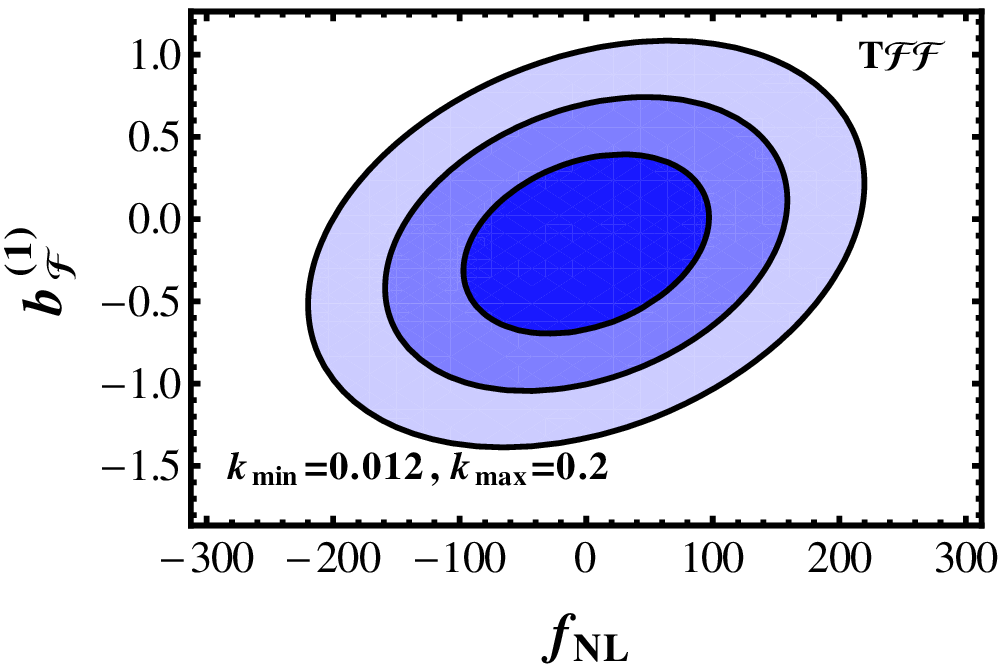}}
\resizebox{140pt}{90pt}{\includegraphics{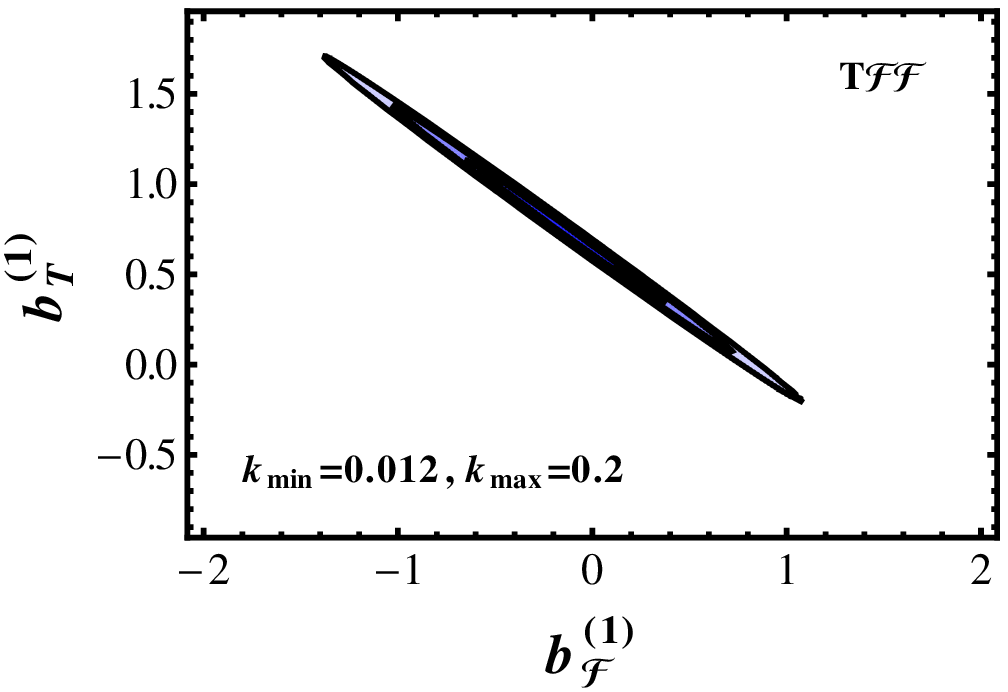}}

\resizebox{140pt}{90pt}{\includegraphics{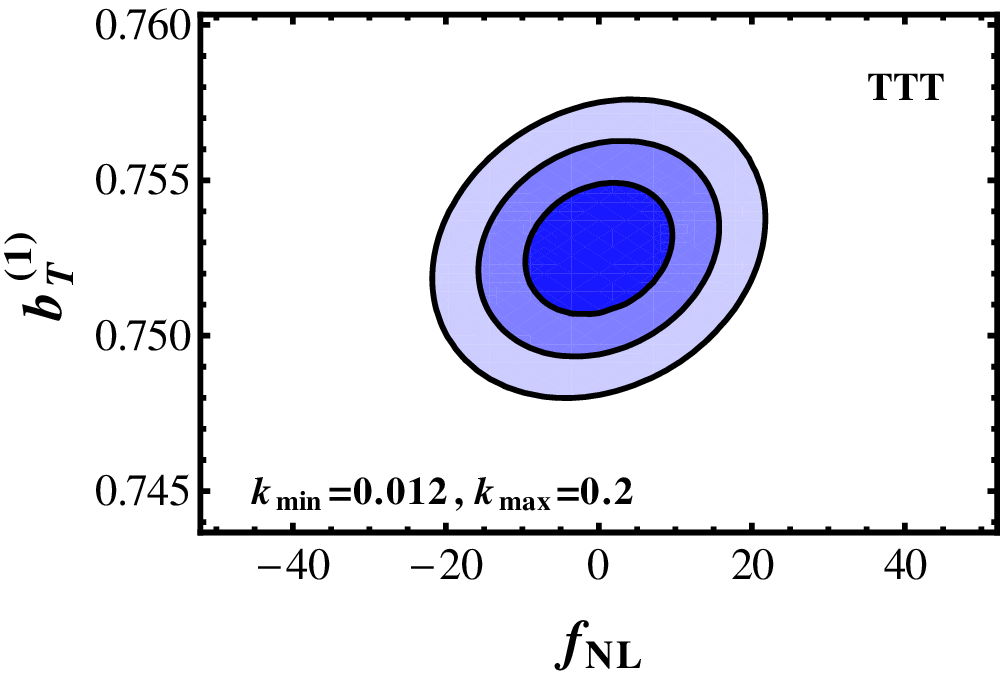}}
\resizebox{140pt}{90pt}{\includegraphics{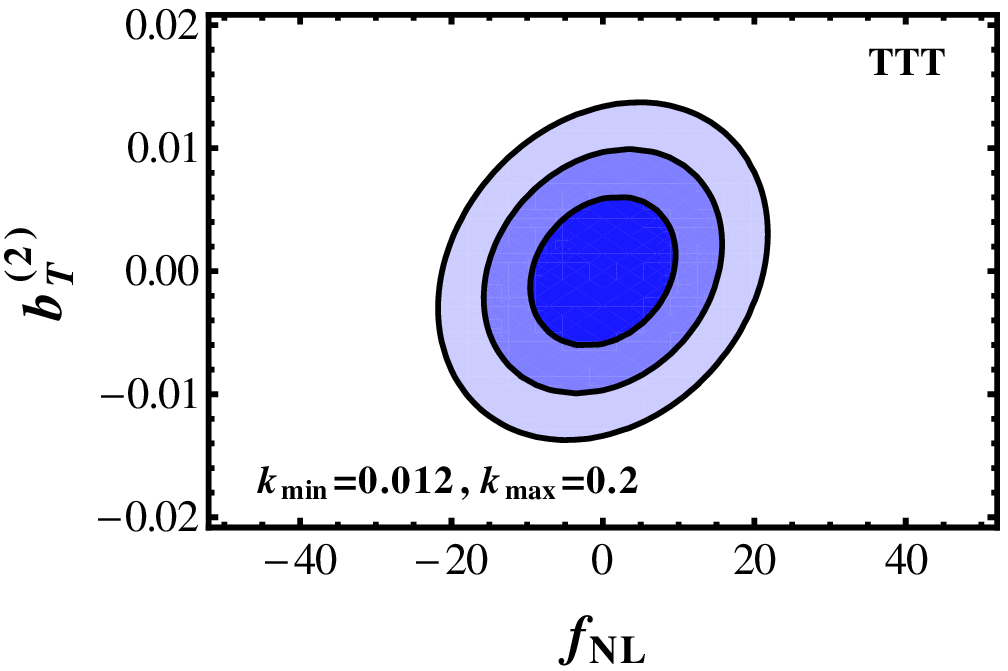}}
\resizebox{140pt}{90pt}{\includegraphics{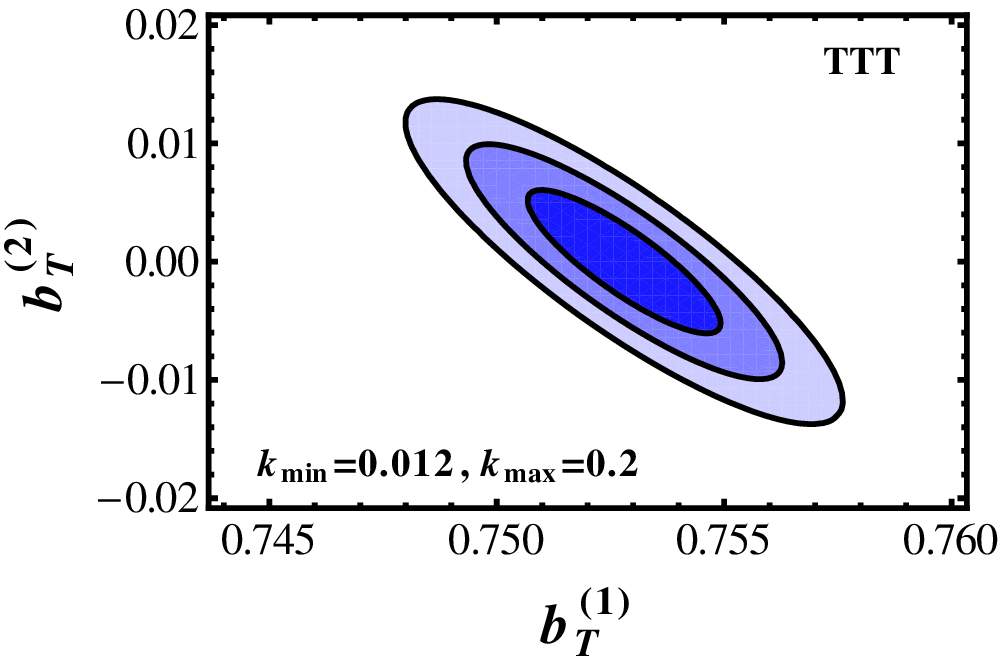}}

\resizebox{140pt}{90pt}{\includegraphics{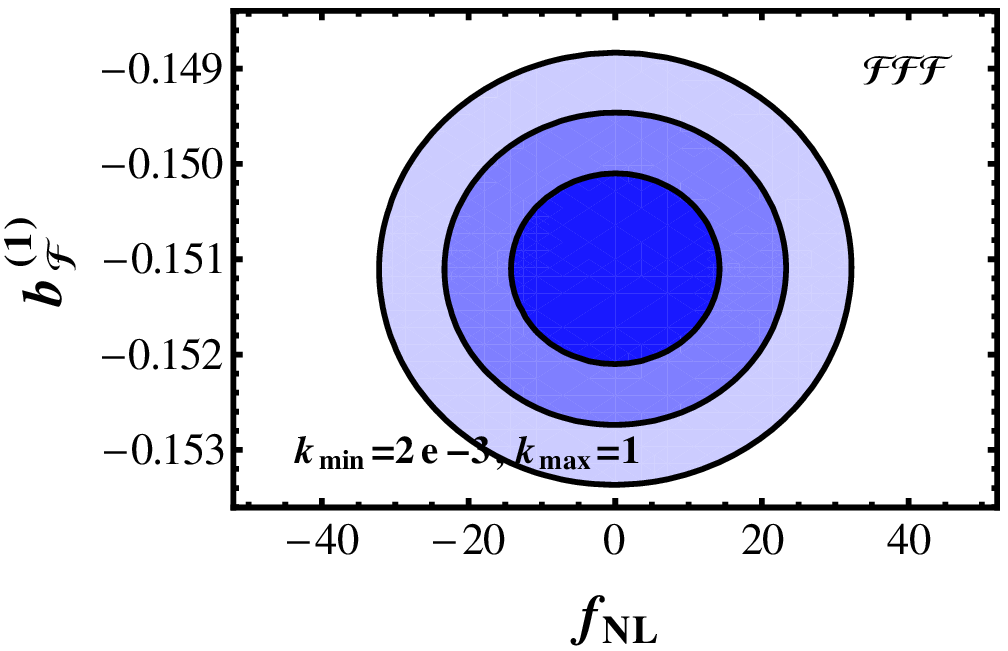}}
\resizebox{140pt}{90pt}{\includegraphics{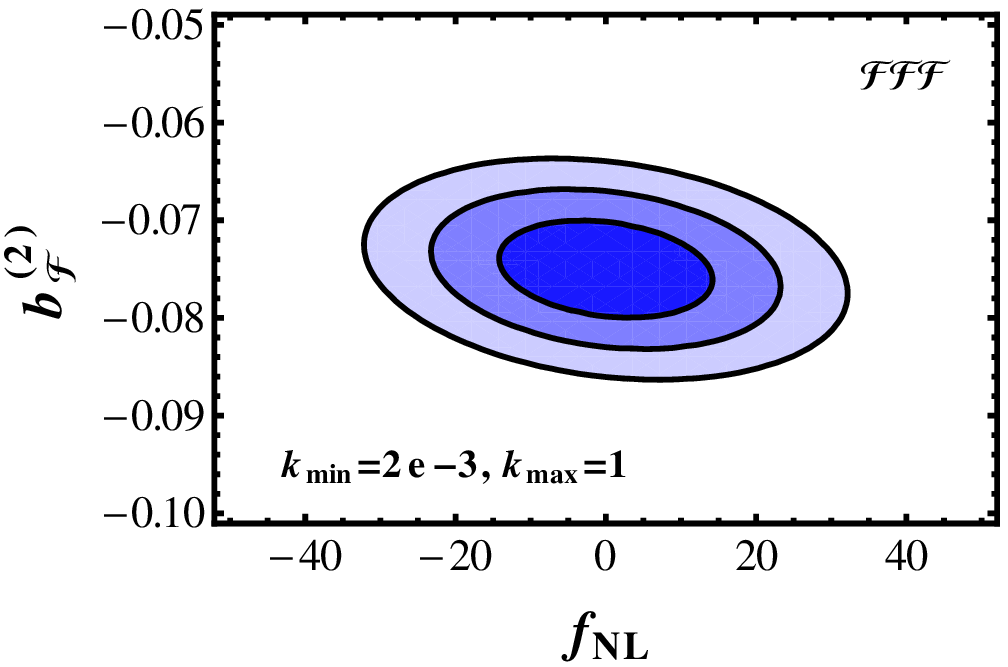}}
\resizebox{140pt}{90pt}{\includegraphics{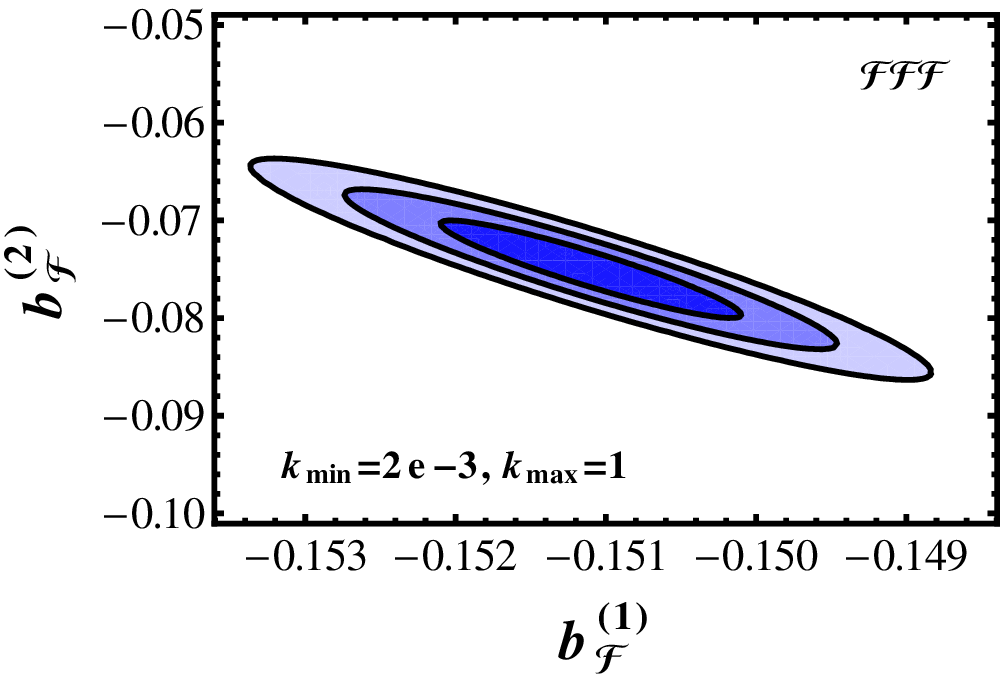}}

\resizebox{140pt}{90pt}{\includegraphics{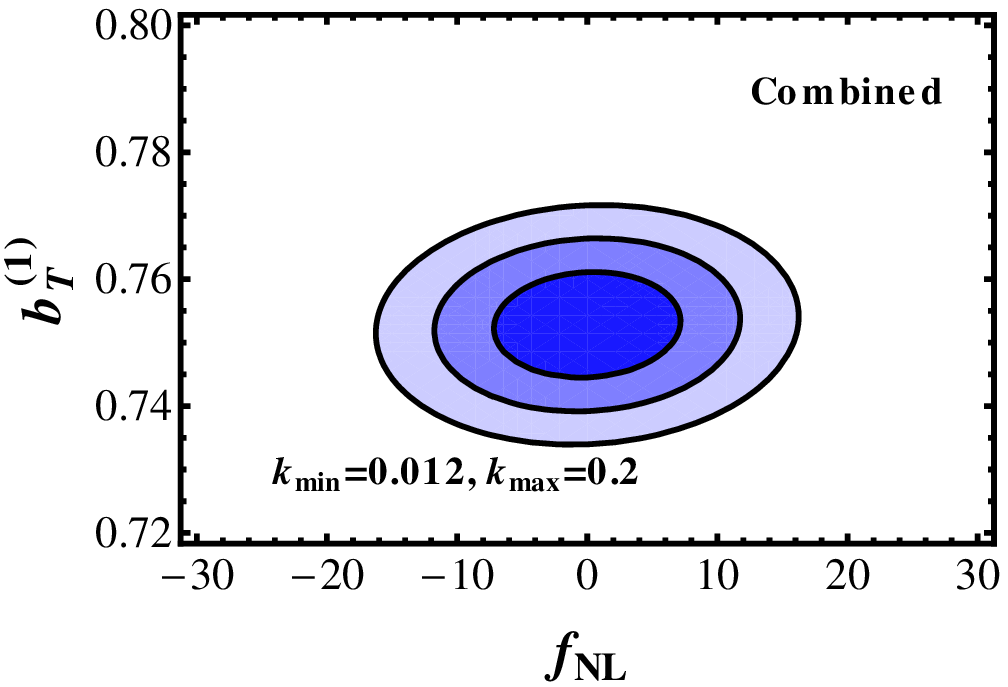}}
\resizebox{140pt}{90pt}{\includegraphics{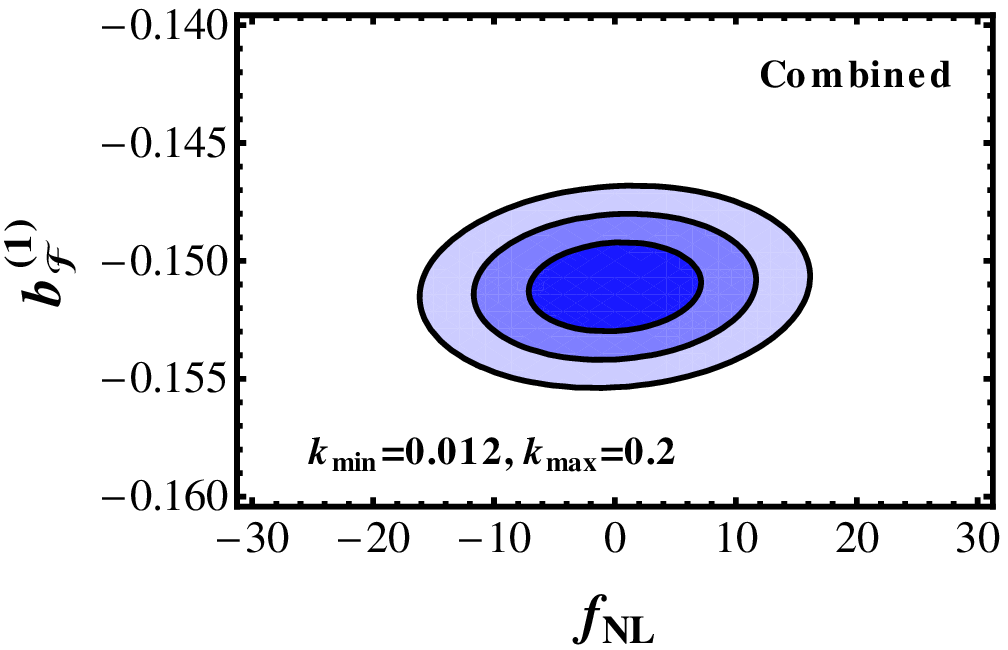}}
\resizebox{140pt}{90pt}{\includegraphics{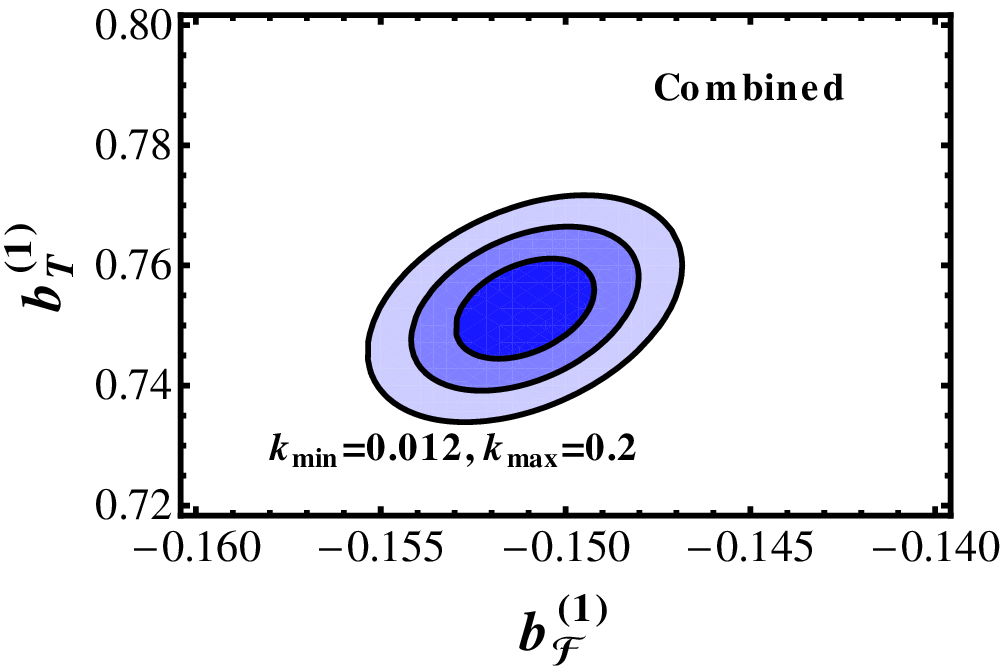}}
\end{center}
\caption{The  $68.3 \%$, $95.4\%$ and $99.8\%$ likelihood confidence 
contours for  $\fnl$ and various bias parameters. Shown in the figure are the 
values $k_{min}$  and $k_{max}$ (unit $\rm Mpc^{-1}$) used to compute the Fisher matrix for different estimators
mentioned in the plots. The bispectrum estimators used for analysis 
are $\F {\rm T T}$, $ {\rm T} \F \F$,  ${\rm T T T} $, $\F \F \F$. The lowest panel 
shows the confidence ellipses for the combined estimator.}
\label{fig:contours}
\end{figure}
We consider a radio interferometric observation of the
redshifted 21-cm signal at a fiducial redshift $z = 2.5$. The choice
of redshift is justified by a apriori knowledge of quasar distribution
which is known to peak between $ z = 2 $ and $ z = 3$. We consider a
futuristic radio observation with a SKA\footnote{http://www.skatelescope.org}
 like radio array with $1400$
antennas, each offering an effective area of $45~{\rm m^2}$ ( fov $\sim
\pi 8.6^2 {\rm deg^{2}}$),  spread out with an approximately uniform
baseline distribution. The observing central frequency is taken to
be $ 406 ~{\rm MHz}$ corresponding to $z= 2.5$. The system temperature is
dominated by the sky temperature and is assumed to be of the form
$T_{sys}= \l[60 \left [ \frac{1+z}{4.73}\right ]^{2.55}+50 \r]~ \rm K$
\cite{santos10}. The noise estimates depend crucially on the volume of
the survey We take $k_{min} = 0.012~{ \rm Mpc^{-1}}$ and $k_{max} = 0.2
~{\rm Mpc^{-1}}$ for a single field of view  observation and consider a total
observing time of $4000~{\rm hrs}$.

Typically, the Ly-$\alpha$ surveys have a much larger sky coverage in
the transverse direction than single field radio observations. In the
radial direction, the frequency bandwidth of radio telescope dictates
the volume of the 21-cm survey. For the Ly-$\alpha$ forest observations
parts of the spectra contaminated by Ly-$\beta$ or $\rm O-VI$ lines at
one end and quasar proximity effect on the other end, may not be  used and
the correlation can me measured in a smaller redshift interval.  The
cross-correlation can however only be measured in the overlapping
regions.  We have considered a Ly-$\alpha$ spectroscopic survey with $
\bar n = 10^{-3}~{\rm Mpc^{-2}}$ which is assumed to be constant across
the redshift range of interest.  We note that $\bar n$ shall in
general be dependent on the central redshift and magnitude limit of
the survey. We also assume that the spectra are measured with a $\rm
S/N = 2$ for $1 \mathring{A}$ pixels. We consider a
 BOSS\footnote{http://cosmology.lbl.gov/BOSS/} like survey
which gives us $k_{min} = 2 \times 10^{-3} ~{\rm Mpc^{-1}}$ and $k_{max}
= 1~{ \rm Mpc^{-1}}$ respectively.  The cross-bispectrum between the
21-cm signal and Ly-$\alpha$ forest may either include two Ly-$\alpha$
component and one 21-cm component or vice-versa. We denote these as
$\cB_{\F \F \rm T}$ and $\cB_{{\rm T T} \F }$ respectively.  Both of
these can be estimated from the same data sets. However, the relative
merits depends on the nature of the observations and quality of the
data.

The cross correlation may be computed in the volume spanned by the
21-cm observation,  which is smaller than the coverage of the
Ly-$\alpha$ survey.  Further, the signal with higher resolution (in
this case Ly-$\alpha$ forest) has  to be smoothed to match the
poorer resolution of the other field. The bispectra are then obtained  for
arbitrary triangles assuming that the primordial power spectrum is 
generated by a quadratic inflaton potential and  the background 
cosmological parameters are frozen to their best fit values 
from  WMAP7 year data~\cite{hazra-2010}. 

We first consider $\cB_{ \F {\rm T T}}$. The Fisher matrix analysis is
used to obtain bounds on the parameters $ (f_{\rm NL}, b^{(1)}_{\rm
  T}, b^{(1)}_{\F})$.  The Likelihood function is assumed to be a
Gaussian, which yields the confidence ellipses in the parameter space
centered at the fiducial values of the corresponding
parameters. Figure~\ref{fig:contours} (top panel) shows these results
for $\cB_{ \F {\rm T T} }$. The tilt of the ellipse measures the
correlation between the parameters, and the semi major/minor axes
measures the maximum uncertainties in measurement of the
parameters. The correlation between parameters $p_i$ and $p_j$ is
measured using a coefficient $r_{i j } = F_{ij}^{-1}/\sqrt{F_{i i
  }^{-1}F_{j j}^{-1}}$.  The strong correlation between $b^{(1)}_{\rm
  T}$ and $ b^{(1)}_{\F}$ ($r \sim 0.997$) reflects the fact that
these parameters occur as a product in the cross-powerspectra. We find
that it is possible to constrain $f_{\rm NL}$ at a level $\Delta
f_{\rm NL} = 17$ for the given observational parameters.  When we
consider $\cB_{\rm T \F \F }$ (see Figure (\ref{fig:contours})) the
bounds on $f_{\rm NL} $ worsens and we have $\Delta f_{\rm NL} =
64$. This is because - as compared to $\cB_{\rm T \F \F }$ the
quantity $\cB_{ \F {\rm T T} }$ has a greater dependence on the 21 cm
signal which appears twice in it and the SKA parameters chosen here
makes the 21-cm signal less noisy as compared to the $2-\sigma$
Ly-$\alpha$ spectra. The situation would be reversed if the
Ly-$\alpha$ forest had high SNR and the 21-cm observation was
relatively more noisy. The numerical results are summarized in the
table (\ref{tab:tab1}).  For the cross bispectra, the volume of the
overlap region dictates the sensitivity. Between the two fields being
used to compute the bispectrum, the one which appears twice should
have higher SNR, to yield better constraints on the parameters.

For the auto bispectra, the survey volume and the observational noise
decides the errors on the parameters being estimated.  For the 21-cm
signal and Ly-$\alpha$ forest auto correlation we choose the
parameters $(f_{\rm NL} , b^{(1)}_{\rm T}, b^{(2)}_{\rm T})$ and
$(f_{\rm NL} , b^{(1)}_{\F}, b^{(2)}_{\F})$ respectively.  We use the
fiducial values $ b^{(2)}_{\rm T} = b^{(2)}_{\F} = 0$ in our Fisher
analysis.  For the Ly-$\alpha$ forest, the SNR for the individual
spectra is low. This is compensated by the large survey volume. With
the given noise parameters, the constraint on $f_{\rm NL}$ from the
21-cm and Ly-$\alpha$ bispectrum are of the same order with bounds $\Delta
f_{\rm NL} = 9.4$ obtained from $\cB_{\F \F \F}$ and $\Delta f_{\rm NL} = 6.3$
obtained from $\cB_{\rm T T T}$. A strong correlation between the
linear and quadratic bias parameters is also seen in both the cases
(see Figure (\ref{fig:contours}) lower two panels). We note here that this is the 
first prediction of constraints on $f_{\rm NL}$ from 21-cm signal and Ly-$\alpha$
 forest using the three dimensional analysis
with the Bispectrum estimator for arbitrary triangular configurations.

\begin{table}[!htb]
\begin{center}
\begin{tabular}{c | c c c c c}
\hline\hline
  & & & &\\
Estimator  &  $\F\F\F$  & ${\rm T}\F\F $ & $\F {\rm T T}$& ${\rm TTT}$ & Combined\\
 \cline{1-1}
Parameters&$p_2=b^{(1)}_{\F}$ & $p_2=b^{(1)}_{\rm T}$&$p_2=b^{(1)}_{\rm T}$ &$p_2=b^{(1)}_{\rm T}$&$p_2=b^{(1)}_{\rm T}$ \\
 &$p_3=b^{(2)}_{\F}$ &$p_3=b^{(1)}_{\F}$ &$p_3=b^{(1)}_{\F}$ &$p_3=b^{(2)}_{\rm T}$& $p_3=b^{(1)}_{\F}$\\
\hline\hline
  & & & &\\
$\Delta\fnl$ & 9.4  &64 &17.2 & 6.3 & 4.7\\
 & & & & \\

$\Delta p_2$ &$7\times 10^{-4}$  &0.28& $3.5\times 10^{-2}$ &$1.4\times 10^{-3}$&$5.5\times 10^{-3}$  \\
 & & & & \\

$\Delta p_3$& $3\times 10^{-3}$&0.36 &$2\times 10^{-2}$&$4\times 10^{-3}$&$1.2\times 10^{-3}$ \\
 & & & & \\
$r_{12}$ &$4\times 10^{-3}$  &-0.31  &0.27 &0.2&$7\times 10^{-2}$ \\
 & & & & \\

$r_{13}$ & -0.21 &0.3 & 0.21& 0.22&0.1\\
 & & & & \\

$r_{23}$&-0.93 & -0.99& 0.99& -0.86&0.4\\
 & & & & \\
\hline\hline
\end{tabular}
\end{center}
\caption{\label{tab:tab1} The bounds on various parameters and their correlations 
obtained from Fisher analysis for different bispectrum estimators.
Note that the parameter $p_1$ denotes $\fnl$ in all the estimators.}
\end{table}

\begin{table}[!htb]
\begin{center}
\begin{tabular}{c | c c c c c}
\hline\hline
  & & & &\\
Estimator  &  $\F\F\F$  & ${\rm T}\F\F $ & $\F {\rm T T}$& ${\rm TTT}$ & Combined\\
 \cline{1-1}
Parameters&$p_2=b^{(1)}_{\F}$ & $p_2=b^{(1)}_{\rm T}$&$p_2=b^{(1)}_{\rm T}$ &$p_2=b^{(1)}_{\rm T}$&$p_2=b^{(1)}_{\rm T}$ \\
 &$p_3=b^{(2)}_{\F}$ &$p_3=b^{(1)}_{\F}$ &$p_3=b^{(1)}_{\F}$ &$p_3=b^{(2)}_{\rm T}$& $p_3=b^{(1)}_{\F}$\\
\hline\hline
  & & & &\\
$\Delta\fnl^{equ}$ & 62  &209 &77 & 37 & 16.8\\
 & & & & \\

$\Delta p_2$ &$8.2\times 10^{-4}$  &0.40& $6.09\times 10^{-2}$ &$1.8\times 10^{-2}$&$5.7\times 10^{-3}$  \\
 & & & & \\

$\Delta p_3$& $4.6\times 10^{-3}$&0.51 &$2.73\times 10^{-2}$&$8\times 10^{-2}$&$1.3\times 10^{-3}$ \\
 & & & & \\
$r_{12}$ &$0.58$  &-0.74  &0.83 &$-0.69$&$-0.26$ \\
 & & & & \\

$r_{13}$ & -0.71 &0.73 & 0.81& 0.88&$-0.38$\\
 & & & & \\

$r_{23}$&-0.95 & -0.99& 0.99& -0.93&0.44\\
 & & & & \\
\hline\hline
\end{tabular}
\end{center}
\caption{\label{tab:tab2} The bounds on various parameters and their correlations 
obtained for the {\it equilateral} template}
\end{table}


\begin{table}[!htb]
\begin{center}
\begin{tabular}{c | c c c c c}
\hline\hline
  & & & &\\
Estimator  &  $\F\F\F$  & ${\rm T}\F\F $ & $\F {\rm T T}$& ${\rm TTT}$ & Combined\\
 \cline{1-1}
Parameters&$p_2=b^{(1)}_{\F}$ & $p_2=b^{(1)}_{\rm T}$&$p_2=b^{(1)}_{\rm T}$ &$p_2=b^{(1)}_{\rm T}$&$p_2=b^{(1)}_{\rm T}$ \\
 &$p_3=b^{(2)}_{\F}$ &$p_3=b^{(1)}_{\F}$ &$p_3=b^{(1)}_{\F}$ &$p_3=b^{(2)}_{\rm T}$& $p_3=b^{(1)}_{\F}$\\
\hline\hline
  & & & &\\
$\Delta\fnl^{orth}$ & 31  &143 &44 &17.6 & 13\\
 & & & & \\

$\Delta p_2$ & $7.8\times 10^{-4}$ &0.37& $5.11\times 10^{-2}$ &$2.2\times 10^{-3}$&$6.3\times 10^{-3}$  \\
 & & & & \\

$\Delta p_3$& $3.5\times 10^{-3}$ &0.48 &$2.41\times 10^{-2}$& $6.1\times 10^{-3}$&$1.6\times 10^{-3}$ \\
 & & & & \\
$r_{12}$ &0.51 &-0.68  &0.74 &$-0.8$ &$-0.49$ \\
 & & & & \\

$r_{13}$ &-0.38 &0.69 & 0.75& 0.79&-0.64\\
 & & & & \\

$r_{23}$&-0.95& -0.99& 0.99&-0.98 &0.57\\
 & & & & \\
\hline\hline
\end{tabular}
\end{center}
\caption{\label{tab:tab3} The bounds on various parameters and their correlations 
obtained for the {\it orthogonal} template.}
\end{table}
\noindent
\begin{figure}[!htb]
\psfrag{kmin}[0][0][1.8]{$k_{min}~( \rm Mpc^{-1})$}
\psfrag{delfnl}[0][0][1.5]{$\Delta\fnl$}
\psfrag{Lymanauto}{~~~~$\F\F\F$}
\psfrag{21auto}{${\rm TTT}$}
\psfrag{LymanLyman21}{~~~~~~~~~~${\rm T} \F\F $}
\psfrag{Lyman2121}{~~~~~$\F{\rm TT}$}
\psfrag{All}{\hskip -45 pt Combined}
\begin{center}
\resizebox{250pt}{180pt}{\includegraphics{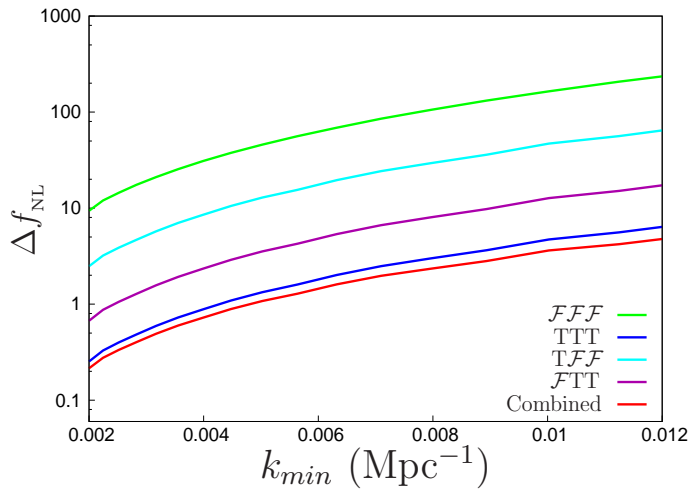}}
\end{center}
\caption{The bounds on $\fnl$ for different estimators as a function of $k_{min}$ assuming 
same SNR and observation time.  Clearly this plot indicates a hierarchy of power to constrain $\fnl$ between 
different estimators. However, this hierarchy is specific to the observational parameters and noise levels associated with the two fields.}
\label{fig:kmindep}
\end{figure}
It may appear from the above analysis that a certain combination of
$\F$ and $T$ has a greater advantage over the others. This hierarchy
arises from the individual noise levels for the respective
observations. One may however combine the information contained in all
the individual estimators to increase the SNR to a maximum for a given
set of observational parameters.  We have performed the full Fisher
analysis using all the $5$ parameters and including all the possible
combinations of $\F$ and $T$. However, the $k-$ range chosen for this
analysis is taken to be corresponding to the common coverage and
resolution of the two surveys. This, as expected yields the best SNR
and a bound on $\Delta f_{\rm NL} \sim 4.7$. Tighter constraints on
the bias parameters are also obtained in the combined analysis (see
table (\ref{tab:tab1})).

Table (\ref{tab:tab2}) and table (\ref{tab:tab3}) shows the results
obtained for the {\it equilateral} and {\it orthogonal} templates respectively.
The degradation of the bounds obtained is on expected lines and we have
$\Delta \fnl^{equ} = 16.8$ and $\Delta \fnl^{orth} = 13$ for the combined estimator.

 As noted earlier, the volume of the survey plays a crucial role in
 the estimation of the parameters.  It fixes the largest scales
 $k_{ min} $ that one may probe. Figure~\ref{fig:kmindep} shows the variation of the
 error on $f_{\rm NL} $ with $k_{min} $ for a given set of
 experimental parameters. It indicates that, for all the bispectra
 estimators considered, there is a monotonic increase of $\Delta
 f_{\rm NL}$ with $k_{min} $. The constraints on $f_{\rm NL}$
 hence, gets worse when $k_{min}$ is increased. This is expected,
 since increasing $k_{min}$ leads to the availability of lesser
 number of k-modes.

In our analysis, the observational parameters have been kept constant
throughout. In reality the relative merit or demerit of a given
estimator will depend on the noise in the data sets. The complete
exploration of the full range of observational parameters for the two
fields is required to be done to judge which estimator yields tighter
constraint on $f_{\rm NL}$ as compared to the others and for what
values of the observational parameters can that be achieved.

The predictions made in our analysis are optimistic and shall
  undergo some degradation with the inclusion of more realistic and
  complete noise or foreground.  However, the present analysis
  indicates that our constraints on $\fnl$ are much better than the SDSS ( $\Delta \fnl \sim 255$ and $\Delta \fnl^{equ} \sim 1775$ )
  and is competitive or better than the  predictions for the CIP \footnote{http://www.cfa.harvard.edu/cip/} survey ( $\Delta \fnl \sim 4.7 $ and $\Delta \fnl^{equ} \sim 51$. See  table 1. in \cite{sefu2007}).

Several observational issues pose severe hindrance towards detecting
the cosmological signals.  For the 21-cm signal, astrophysical
foregrounds which are several order higher than the signal completely
submerge the cosmological information \cite{fg1}. Several methods using the
distinct spectral behavior of the foregrounds as compared to the
signal has been proposed to clean the observed maps and retrieve the
cosmological signal \cite{fg2}. In this work we have assumed that for the 21-cm
field, foreground cleaning has been done.  For the Ly-$\alpha$ forest,
a host of issues need to be addressed. This includes a
proper modeling and subtraction of the continuum, tackling of contamination from
metal lines in the forest \cite{viel2004}, effect of Galactic super winds, to mention
just a few.  Redshift space distortion plays a crucial role on scales
of our interest and is required to be incorporated in our analysis.
We intend to take this up along with a detailed analysis of
observational aspects and instrumental noise in a future work.

We conclude by noting that the three dimensional distribution of the
post-reionization neutral hydrogen as probed using the 21-cm or the
Ly-$\alpha$ forest, may be potentially used to estimate various
auto/cross bispectra and thereby enrich our understanding of the early
Universe.

\section{Acknowledgements}

D.K.H wishes to acknowledge support from the Korea Ministry of
Education, Science and Technology, Gyeongsangbuk-Do and Pohang City
for Independent Junior Research Groups at the Asia Pacific Center for
Theoretical Physics.


\end{document}